\documentclass{article}
\usepackage{mystyle}
\usepackage{epsfig}

\begin{document}
\setcounter{totalnumber}{10}

\title{On the foundations of statistical mechanics:\\
ergodicity, many degrees of freedom and inference}

\author{Sergio Chibbaro$^{1,6}$ and Lamberto Rondoni$^{2,3,6}$ and Angelo Vulpiani$^{4,5,6}$\vspace{2mm} \\
$^{1}$ Sorbonne Universit\'es, UPMC Univ Paris 06, \\
CNRS, UMR7190, Institut Jean Le Rond d'Alembert, \\
F-75005 Paris, (France) \\
$^{2}$  Dip. Scienze Matematiche, Politecnico di Torino, \\
C. Duca degli Abruzzi 24, Torino. \\
$^3$ INFN, Sezione di Torino, Via P. Giuria 1, Torino (Italy). \\
$^{4}$ Dipartimento di Fisica,  Universit\`a"La Sapienza" Roma (Italy)\\ 
$^5$ ISC-CNR, Piazzale A. Moro 2 - 00185 Roma (Italy) \\
$^6$ Kavli Institute for Theoretical Physics China, CAS, Beijing 100190, China
}
\maketitle

\section{Introduction}
Statistical mechanics has been  founded by Maxwell, Boltzmann and Gibbs who
intended to describe in terms of microscopic interactions the properties of 
macroscopic objects, {\em i.e.} systems made of so many particles that 
a statistical treatment is required.
For almost all practical purposes one can say that the whole subject of
statistical mechanics consists in evaluating a few suitable functions
and quantities, such as the partition function, the free energy and the
correlation functions.

Since the discovery of deterministic chaos it is plain that statistics (e.g.\ over
ensembles of objects following the same dynamical rules, or over time series of 
fluctuating quantities) is 
unavoidable and useful even in systems with a few degrees of freedom.
On the other hand, no universal agreement has been reached so far about the 
fundamental ingredients for the validity of statistical mechanics \cite{Skl_95,EmL_02}.

The wide spectrum of positions includes Landau's and Khinchin's belief that the 
main ingredient in producing the common macroscopic behaviours is the vast number of
microscopic degrees of freedom, which makes (almost) completely irrelevant details
of the microscopic dynamics such as the validity of the standard notion of ergodicity,
developed in the mathematical literature. But it includes also the opinions of those,
like Prigogine and his school, who consider chaos, in the sense of the positivity of
Lyapunov exponents, as a fundamental ingredient \cite{Pri_84}, although this (to us
untenable) position has been strongly criticized and dismissed in numerous essays, 
{\em e.g.} Ref.\cite{bricmont1995science}.

%For almost all practical purposes one can say that the whole subject of
%statistical mechanics consists in the evaluation of a few suitable
%quantities (for example, the partition function, free energy,
%correlation functions).

On the other hand, for decades, the ergodic problem has been often completely neglected and 
the (so-called) Gibbs {\em ensemble} approach has been accepted on the grounds that ``it works''. 
Moreover, this point of view has been commonly associated with supposedly
universally acceptable principles, such as the maximum entropy principle (MEP), according to 
which, in good substance, all the statistical properties of a given object can be inferred 
simply by maximazing a kind of information theoretic entropy, under the constraints that are
known to act on the system itself, \cite{MEP_2013}.
But also this approach is hardly satisfactory, especially if one thinks that
understanding the mechanisms underlying macroscopic physics is not less
important than computing useful quantities.  

The present paper is meant to give
a simple introduction to the problem of the connection between microscopic dynamics
and statistical laws. For sake of simplicity, we mostly refer to non-dissipative dynamics,  
since dissipation adds technical difficulties to the conceptual issues, although part of our 
discussion extends beyond this limit. In particular, the relevance of chaos and ergodicity is here
confronted with that
of the large number of degrees of freedom. In Section 2, we review the microscopic connection, 
along the lines of Boltzmann's approach, and of its further developments. In Section 3, we discuss 
the falsifiability of statistical mechanics and its role as statistical inference. In Secion 4, we 
draw our conclusions.

\section{From microscopic to macroscopic}

Once the atomistic view is accepted, the identification of the issues that may benefit from it still poses a challenge. No doubt, matter is made of atoms, and what happens 
to any piece of matter depends on what happens to its atoms; but which aspects of the 
behaviour of that object may be understood in terms of what we know about atoms? Which 
phenomena may be elucidated by the laws which we use to describe the behaviour of 
atoms? The fact is that our description of the atomic world is a direct extension of 
classical (Newtonian) mechanics,\footnote{Conceptually, quantum mechanics does not change 
the picture \cite{lebowitz1993boltzmann}.} which has been very {successfully} developed to describe 
systems made of a small number of objects, while the properties of matter result from 
the cooperation of exceedingly large numbers of atoms. 
At the fundamental level the main problem of statistical mechanics
is to establish a link between the thermodynamics description
and the dynamics ruling the time evolution of the system.
Some books devoted to foundational aspects of statistical mechanics 
are the following: \cite{Skl_95,EmL_02,castiglione2008chaos}.

%At variance with quantum mechanics, whose foundations 
%have attracted a plethora of works both on the philosophical and 
%physical side, 
%is not a very popular
%topics. In order 

To treat the main conceptual problems of statistical physics, let us start from classical mechanics,
considering the space $\phasespace$ of all microscopic configurations of a given system of interest, known 
as the \emph{phase space}. In the 
case of the simple monatomic gases, $\phasespace$ may be taken to be the 6$N$ dimensional phase space $\zR^{6N}$, 
i.e.\ the space of all coordinates and momenta $\Gamma = (\qq,\ppp)$ of the 
$N$ atoms. {If the elementary constituents of the system cannot be approximated by 
point particles, but still obey classical mechanics, one would have to refer to a larger
phase space, which has more dimensions, but the basic idea does not change.} 
Given {a vector} 
$\zG \in \phasespace$ { representing an initial microstate}, the state at a later time may be denoted by the evolved vector 
$S^t \zG \in \phasespace$, corresponding to the changes in positions and velocities of the 
atoms, where $S^t$ is the evolution operator for a time $t$.

In the case of an isolated system of particles, this evolution is
expressed by Hamilton{'s} equations of motion, which can be compactly written as:
\begin{equation}
\dot{\qq} = \frac{\partial H}{\partial \ppp}~; \qquad \dot{\ppp} = 
-\frac{\partial H}{\partial \qq}
\label{HamEq}
\end{equation}
where $H$ is the Hamiltonian of the system, and $H(\qq,\ppp)=E$ determines the energy surface
in which the time evolution $S^t \zG$ is confined for all $t$.
Then, one assumes that each microstate $\Gamma \in \phasespace$ implies a particular 
value $\observable(\zG){\in\zR}$ for each observable quantity $\observable$. 

Macroscopic measurements do not occur instantaneously, but take a certain amount of time: as a consequence 
of the very rapid and perpetual motion of the atoms, they do not yield a precise value $\observable(\zG)$, 
because $\Gamma$ changes in that time.  Instead, a
measurement yields the average of the values that $\observable(S^t \zG)$ takes 
while $S^t \zG$ explores the phase space. For instance, the pressure of a gas measured by a 
manometer is determined by the average of the variations of the molecular momenta, which 
occur when the molecules hit the surface of the sensor.

\subsection{Boltzmann's grand vision}
A mathematical expression for a macroscopic measurement, i.e.\ for the average over 
the myriad microscopic events performed by our senses or by our measurement tools,
is given by:
\begin{equation}
\overline{\observable}^T(\Gamma) =
\frac{1}{T} \int_0^T \observable(S^t \Gamma)~ d t~,
\end{equation}
where $T$ is the duration of the measurement and $\zG$ is the initial microstate. 

In general, the result of a measurement could depend on both $T$ and on $\zG$. 
The dependence on $T$ makes the result of the measurement{ subjective}, as $T$ may be 
varied at will by the experimentalist, while the dependence on $\zG$ makes it stochastic, 
because the initial microstate can neither be controlled nor identified. 
For a scientific theory to be conceived, these dependencies must be kept under control. 

In particular, for the connection with thermodynamics, the microscopic events must
occur on time scales much shorter than the observation scales. If that is the case, 
one {\it may hope} that the quantity $\mathcal{O}(S^t \zG)$ explores the interval of 
all its possible values so rapidly that the time required by a measurement makes
$\overline{\observable}^T(\Gamma)$ indistinguishable from its asymptotic value
almost independently of $\zG$, washing away subjectivity and stochasticity.
The appropriate value of $T$ depends on the sensitivity of the measurement tool: it is 
shorter for higher sensitivity, but it nevertheless remains virtually infinite with 
respect to molecular characteristic times, such as the average time between two molecular collisions. 
Therefore, a theory of measurements relies on the existence of the following limit: 
\begin{equation}
\overline{\observable}(\zG) =
\lim_{T\rightarrow\infty} \frac{1}{T} \int_0^T \observable(S^t \Gamma)~ d t  
\label{Taverage}
\end{equation} 
and on the irrelevance of $\zG$ for its value. Will that ever be the case for given microscopic
dynamics?

The existence of the time limits was
proven by Birkhoff in 1931 for a wide class of systems, while the dependence
on $\zG$ poses many more problems, from a mathematical perspective. Indeed, if thermodynamics
applies, the properties of a macroscopic system can be predicted without any knowledge 
of $\zG$. This is essentially the content of {Boltzmann's celebrated}
\emph{ergodic hypothesis}. Roughly speaking, in its modern standard formulation,
this hypothesis states that each hyper-surface with constant energy is completely accessible to every motion with the given energy.
As a corollary, for each trajectory, the residence-time in any region is proportional to the volume
of the region.

When the ergodic hypothesis holds, $\mathcal{M}$ may be endowed with a probability density
$\rho$ reflecting the frequency with which the different regions of $\mathcal{M}$ are visited
during the evolution. This probability distribution, also called {\em ensemble} 
\cite{Fer_56},  associates 
higher weights to the regions which are visited more often, and lower weights to those which
are visited less frequently
so that: 
\begin{equation}
\overline{\observable}(\Gamma) = \int_\phasespace \observable(\zG)  \, \rho (\zG){\rm d} \zG
\equiv \langle \observable \rangle_{\rho}
\label{EH}
\end{equation}
for all initial conditions $\Gamma \in \phasespace$, except a set of vanishing probability. 
The quantity $\langle \observable \rangle_{\rho}$ is called the  phase average, and a given
system is called {\em ergodic} with respect to the probability distribution $\rho{\rm d} \zG$
if the time averages of {\em all} its observables obey Eq.(\ref{EH}).\footnote{More precisely,
given an invariant probability measure $\mu$ on $\phasespace$, the statement must hold for all 
functions $\observable$ which constitute a suitable function space, e.g.\ Birkhoff chose
$L^1(\phasespace,\mu)$. Also, note that the equality 
$\overline{\observable}(\Gamma)=\langle \observable \rangle_{\mu}=
\int_\phasespace \observable(\zG)  \, \mu({\rm d} \zG)$, for $\mu$-almost-all $\zG$, defines ergodicity
in full generality, hence even for the singular measures $\mu$ required to describe steady states
of dissipative dynamics.}

To explain the physical meaning of Eq.(\ref{EH}), in 
his well known  book on thermodynamics, Fermi says that: {\it Studying the 
thermodynamical state of a homogeneous fluid of given volume at a given temperature 
(the pressure is then defined by the equation of state), we observe 
that there is an infinite number of states of molecular motion that correspond to it. With 
increasing time, the system exists successively in all these dynamical states that correspond 
to the given thermodynamical state. From this point of view we may say that a thermodynamical 
state is the ensemble of all the dynamical states through which, as a result of the 
molecular motion, the system is rapidly passing.}\footnote{Fermi's statement must not be interpreted 
literally: it must be intended, as above, in terms of the values that the observables 
of interest can take, and not in terms of trajectories $\{ S^t \zG \}$ densely exploring the 
phase space. This would require completely unphysical exceedingly long time scales.}

The practical importance of the ergodic hypothesis is evident: knowledge of the initial 
microstate $\zG$ and of the solution of Hamilton's equations, which yields $S^t \zG$,
are unnecessary if the system is ergodic.

{Given all this}, a question arises: how do the averages of the microscopic properties 
of a system of particles relate to the thermodynamic properties of the object {composed of} those particles? The first, in fact, consist of a myriad of mechanical parameters, such 
as the particles' energy and momentum, while the others consist of just a few measurable 
quantities, like temperature and pressure. 
%The corresponding diversity of fundamental 
%terminology, qualifies the reduction of thermodynamics to mechanics as ``heterogeneous'' 
%reduction \cite{Nag_79}, a condition which may prevent the logical derivation of the 
%former theory from the latter. As recalled in the last chapter, for that to be possible, Nagel requires the existence of relations 
%between the terms of the theory to be reduced and elements of the vocabulary of the reducing 
%theory. Such bridge laws must reflect 
%a kind of identity between the objects of study of the two theories, 
%and must be empirically supported.

The conceptual (and technical)  law which associates thermodynamics with the classical 
mechanics of atoms was proposed by Boltzmann and it is engraved in his tombstone:
\be
S = k \log W~.
\label{klogw}
\ee
This celebrated relation connects the thermodynamic entropy $S$ of an object in the macroscopic 
state $X$, to the volume $W$ of all microstates in $\cal M$ which correspond to the same $X$. 
For example, considering the macrostate $X$ corresponding to a given energy $E$, 
one typically considers the energy shell $E - \zd E \le H(\qq,\ppp) \le E$, with small 
$\zd E$, and obtains:
$$
W = \int_{E - \zd E \le H(\qq,\ppp) \le E} {\rm d} \qq {\rm d} \ppp~.
$$
The microcanonical probability distribution is constant in the energy shell so that:
\begin{displaymath}
 \rho (\zG) = \left\{ \begin{array}{ll}
\frac{1}{W} & \; \textrm{if  $ E - \zd E \le H(\qq,\ppp) \le E$} \\
0 & \textrm{otherwise}
\end{array} \right.
\end{displaymath}
In the terms of philosophy of science, Equation (\ref{klogw}) qualifies as a bridge law\cite{Nag_79}, 
because $S$ is a thermodynamic quantity, while $W$ is a microscopic entity. 
Once it has been introduced, further mechanical properties of our description 
of the microscopic dynamics may be related to as many other thermodynamic quantities, thus
bridging the gap between micro- and macro-descriptions. In particular, in the {\em entropy 
representation} of thermodynamics, one obtains the temperature as:
\be
\frac{1}{T} = \frac{\partial S}{\partial E}~, \qquad
\ee
and the free energy as $F = E - TS$, where $E$ is the mechanical energy of the system.
Note that some authors, including Nagel\cite{Nag_79}, take
the relation between the temperature $T$ and the average kinetic energy $K$ as
the bridge law. This is incorrect, as thermodynamics requires a thermodynamic potential, such 
as the free energy, which is a function of the other relevant thermodynamic variables.
In addition, the expression of $T$ in terms of $K$, which had already been guessed by Bernoulli, 
holds only for a special class of phenomena.

The bridge law (\ref{klogw}) has important consequences supported by empirical evidence, 
including, in particular, those derived from Einstein's version:
\be
P(\alpha) \sim e^{(S(\alpha)-S_{\rm eq})/k}
\label{ARver}
\ee 
where, apart from the normalization constant, $W$ has turned into the probability $P(\alpha)$ of the collection 
of microstates which correspond to the same value for the macroscopic variable $\alpha$, $S_{\rm eq}$ is the equilibrium 
entropy and $S(\alpha)$ is the entropy of a spontaneous fluctuation of $\alpha$, produced by the cooperation
of many microscopic motions. Formula (\ref{ARver}) is meant to 
{represent} the probability of fluctuations about equilibrium states of microscopic mechanical 
quantities, such as the energy $E$ of a system in contact with a heat bath at temperature 
$T$. As these fluctuations are related to observable quantities, they can be {characterised} by 
macroscopic equilibrium experiments. For instance, if {$\alpha$} is the energy, one identifies the average $\langle E \rangle$ 
of the energy with the internal energy $U$ of the system, and {introduces} the standard 
deviation $\sqrt{\langle E^2 \rangle - \langle E \rangle^2}$, which measures the size 
of the fluctuations. We obtain:
\be
\langle E^2 \rangle - \langle E \rangle^2 = k T^2 C_v
\label{EneCv}
\ee
where $C_v$ is the heat capacity at constant volume. Because $C_v$ is extensive, hence 
proportional to the number $N$ of particles in the system, the relative
size of the energy fluctuations is negligible in large systems:
\be
\frac{\sqrt{\langle E^2 \rangle - \langle E \rangle^2}}{\langle E \rangle} 
\sim O \left( \frac{1}{\sqrt{N}} \right) \to 0~, \quad \mbox{for }~ N \to \infty
\label{EneFlu}
\ee
In other words, the fluctuations of the microscopic mechanical quantity $E$ grow with the
system size proportionally to $\sqrt{N}$, but are negligible with respect to the observable 
internal energy $\langle E \rangle = U$, which is of order $O(N)$. This should not lead {one }to 
relegate fluctuations {to the set of only} marginally interesting phenomena. Indeed, in his 
search of an ultimate proof of the existence of atoms, Einstein {realised} that Eq.(\ref{EneCv}): 
\begin{itemize}
\item[]
{\em would yield an exact derivation of the universal constant {\em [$k$ or, 
equivalently, Avogadro's number $N_A$]} if it were possible to determine the average of the
square of the energy fluctuations of the system.}
\end{itemize}
He successfully applied this idea to describe the Brownian motion\cite{Ein_56}.

\subsection{Beyond the mathematical limitations of ergodic theory}
The effort to turn the physical notion of measurement into an appropriate mathematical 
one has led to the issue of ergodicity, which seems to cleverly frame the connection 
between mechanical and thermodynamical quantities. Unfortunately,  
apparently modest consideration{ of} real-life systems, such as the insensitivity 
of thermodynamic quantities to microscopic states, {raises} deep mathematical questions.
In particular, 
%For instance, one considers a system comprising $N$ classical particles in $d$ dimensions, 
%whose dynamics is formally described by the following equations:
%\begin{equation}
%\dot{\Gamma} 
%= G (\Gamma)~; \quad  \Gamma = (\qq,\ppp)  \in \phasespace \subset \zR^{2dN} ,
%\label{xdot}
%\end{equation}
%where $\phasespace$ is the phase space, and the vector field $G$ is determined by the 
%forces acting on 
%the system and by the particles' interactions.  
%Furthermore, (\ref{Taverage}) is not particularly suitable for further theoretical derivations, 
%given its highly implicit form. 
one faces the problem of identifying the probability density $\rho$ {that} describes systems in 
equilibrium, or evolving towards equilibrium,
and the fact that requiring Eq.(\ref{EH}) to hold for {\em all} 
phase functions is too demanding, compared to the needs of thermodynamics.\footnote{We state that a physical 
system is in {\em equilibrium}  
if all currents -- of mass, momentum, heat, etc. -- vanish, and the system is uniquely described by 
a (typically quite small) set of \emph{state variables} which do not change with time.} 
In addition, the time scales over which the ergodicity of a system of many degrees of 
freedom would be obtained are astronomically larger than the physically relevant time 
scales.

The celebrated work by Fermi, Pasta and Ulam, concerning a chain of nonlinear 
oscillators, further showed that ergodicity may be violated even by the simplest 
particles systems.
{Indeed, in their numerical simulations, known as the FPU
experiment, Fermi and coworkers showed that a typical Hamiltonian system is not ergodic.
This fact was totally unexpected, at that time, and was only later explained in the 
sophisticated mathematical terms of KAM theory~{\cite{Cen_09}}.} 

Quite surprisingly, the very simple probability
distributions known as {\em microcanonical, canonical} and {\em grand-canonical} 
ensembles describe very well most equilibrium situations. When legitimate, this is an 
extremely powerful way of proceeding, whose success seems to rest
on our limited knowledge of the microscopic dynamics. 
{However one should be {wary} of possible 
misunderstandings. In particular, ensembles are often described as {\em fictitious collections 
of macroscopically identical copies of the object of interest, whose micro{}states differ 
from each other.} While this maybe a convenient perspective, one should not
forget that their purpose is to describe the properties of a single system, whose 
microstate evolves forever. We can say that the word ``statistical ensemble'' is nothing but a way
to indicate the probability density of $\Gamma$.}

But why? The foundations of the ergodic hypothesis  
look shaky, and its success puzzling, if no further explanation is given.
To address these issues, Khinchin pioneered 
an approach based on the following premises\cite{Khi_49}: 
\begin{itemize}
\item[{\bf a)}] statistical mechanics concerns systems with a large number of degrees of 
freedom; 
\item[{\bf b)}] the physical observables are but a few and quite special functions; 
\item[{\bf c)}] it is physically acceptable that ensemble averages do not coincide with 
time averages, on a small set of phase space trajectories;
\end{itemize}
As appropriate for rarefied systems, he considered dynamics whose Hamiltonians are the 
sum of single particle contributions:
$$
H=\sum_{n=1}^N H_n(\qq_n,\ppp_n) 
$$
and restricted the space of observables to the \emph{sum functions}{ -- }functions 
defined as sums of single particle contributions $f_n$:
$$
f(\zG)=\sum_{n=1}^Nf_n(\qq_n, \ppp_n)~.
$$
The pressure and the kinetic energy are examples of such functions. Then,
denoting by $\langle \, . \, \rangle$ the microcanonical ensemble average, Khinchin 
demonstrated that:
$$
\textrm{Prob} \left( \frac{ |\overline{f} - \langle f \rangle|}
 {|\langle f \rangle|} \ge K_1 N^{-1/4} \right) \le K_2 N^{-1/4}~,
$$
where $K_1$ and $K_2$ are constants. This means that the microcanonical averages of sum 
functions  differ from their time averages by more than a (small) relative tolerance only
along a set of trajectories whose probability vanishes in the $N \to \infty$ limit. 
The problem is that the initial conditions of the microstate must be taken within a proper
subset of the phase space, but Khinchin showed that the fraction of volume of phase 
space which lies outside this subset vanishes in the $N \to \infty$ limit.

Ultimately, from various standpoints, Khinchin's theory ascribes the good statistical properties 
required for {}normal thermodynamic behaviour to the fact that $N$ is very large. 
From his perspective, the details of the microscopic dynamics appear practically irrelevant for 
the physics of rarefied gases. An important extension of this approach, which goes beyond the 
low density gas, was obtained by Mazur and van der Linden, \cite{Maz_63}. These authors did not require 
the Hamiltonian to be separable in single particle contributions, but admitted particles to
interact only through short-range interaction potentials and, like Khinchin, considered only 
sum variables. They proved that their systems can be treated as consisting of many 
non-interacting parts. 

Although even this theory is not completely satisfactory, because the set of sum variables 
is too limited for dense systems, the works of Khinchin and of Mazur and van der Linden 
clarify why one should not be surprised that the ergodic hypothesis applies so generally 
in physics: macroscopic systems are made of very many particles.

\section{Remarks}
Let us conclude with some provocative remarks and a brief summary.
\subsection{Is Statistical Mechanics falsifiable?}
The question may seem provocative but deserves a brief excursion.

In statistical mechanics tests, the typical excercise that students have to solve is the following: 
given a box of volume $V$ containing $N$ particles interacting according to
a known potential $U(|{\bf q}_i -{\bf q}_j|)$,
assuming thermal equilibrium at temperature $T$,
is it possible to calculate specific heat, state equation, etc.?
For instance, for dilute gases it is possible to write the virial expansion
\begin{equation}
{p \over k_B T}= \rho + b_2(T)\rho^2+ b_3(T)\rho^3+ {\ldots}
\end{equation}
where $\rho = N/V$ and virial coefficients $b_1,b_2, {\ldots}$ can be expressed in terms
of $U(r)$. 
Calculations should be then compared against experiments.

In practice, however, this approach cannot be pursued. 
Even if the dynamics is classical, the potential $U(r)$ has quantum origins and is not known.
It should be computed from the Schr\"odinger equation, but a good approximation can only be obtained 
in a few lucky cases. In general, 
it is necessary to tackle the problem in a very different way.

First of all, one chooses a given form for $U(r)$, for instance, for simple liquids the Lennard-Jones 
potential is very popular:
\begin{equation}
U(r)= 4 \epsilon \Bigl[ \Bigl( {\sigma \over r} \Bigr)^{12} 
- \Bigl( {\sigma \over r} \Bigr)^{6}  \Bigr] \,\, ,
\end{equation}
where $\epsilon$ and $\sigma$ are two parameters.
Then, one computes the virial coefficients $b_2(T), b_3(T), ...$ as functions of $U(r)$, i.e.\
of the parameters $\epsilon$ and $\sigma$ and, finally, $\epsilon$ and $\sigma$ are determined
by comparison with the experimental data \cite{Att_02}.

Since this procedure is self-consistent, one could conclude that statistical mechanics cannot be falsified. 
While this discussion sheds light on the complex relation between 
microscopic and macroscopic worlds, the conclusion is incorrect. Indeed,
statistical mechanics predicts many other non-trivial properties that can 
be checked experimentally. Among the others, the Maxwell-Boltzmann distribution
for the molecules, phase transitions and universality of critical phenomena\cite{Ma_85,Peliti}.
Interestingly, in all these cases, the knowledge of the potential $U(r)$ is not needed: the
Maxwell-Boltzmann distribution is completely independent of $U(r)$,
whereas critical phenomena are influenced only by some qualitative properties of this potential,
which delimit the universality class.

\subsection{Statistical mechanics as statistical inference?}
According to a radically anti-dynamical point of view, statistical mechanics is 
but a form of statistical inference rather than a theory of objective physical 
reality. Under this light, probabilities measure the degree of truth of a logical 
proposition, rather than describing the state of a system.

In this context, Jaynes proposed the MEP as a general 
rule {for finding} the probability of a given event when only partial information is 
available. If the mean values of $m$ independent functions $f_i({\bf x})$ are 
given:
$$
c_i= \langle f_i \rangle = 
\int f_i({\bf x})\rho({\bf x}) d {\bf x} \,\,\,\, i=1, ... , m \,\,\, ,
$$
the MEP rule determines the probability density $\rho$ of the events compatible with 
these mean values, by maxim{is}ing the ``entropy'' 
$$
H= - \int \rho({\bf x}) \ln \rho({\bf x}) d{\bf x} \,\, ,
$$
under the constraints $c_i= \langle f_i \rangle$. Using the
Lagrange multipliers one easily obtains
$$
\rho({\bf x})= 
{1 \over Z} \exp \sum_{i=1}^m \lambda_i f_i({\bf x})
$$
where $\lambda_1, \lambda_2 ...\lambda_m$ depend on $c_1, c_2,  ... , c_m$.
For instance, for systems with a fixed number of particles subjected to the unique 
constraint that their mean energy is fixed, 
the MEP leads to the canonical distribution in a very simple fashion.

As a technical but rather important detail{,} we note that this holds only 
if ${\bf x}$ is the vector of the canonical coordinates (i.e.\ positions and momenta 
of the particles). Analogously, for systems of varying numbers of particles, the grand 
canonical distribution is obtained {by }additionally constraining the mean {number of} particles.  
Many find in these facts an unquestionable proof of the validity of the MEP, but we will see
that it is just a matter of fortunate coincidence, related to the choice of canonical 
coordinates.

The most frequent objection to the MEP is summarised by the {maxim} {\it Ex
nihilo nihil}, or citing the title of one of Peres's papers, ``Unperformed experiments have no 
results''~\cite{Per_78}, i.e.\ our ignorance cannot be credited for inferences about real phenomena,
\cite{Ma_85}. In spite of the optimistic claims of the MEP enthusiasts,
to the best of our knowledge MEP has only produced different, sometime more elegant, derivations 
of previously known results\cite{Peliti}.
Apart from this very general observation, 
the weakest technical aspect of the MEP approach is the dependence of the results
on the choice of the variables.

For {simplicity's sake}, consider a scalar random variable $X$, ranging over a 
continuum, whose probability distribution function is $p_X$. 
It is easy to real{is}e that the ``entropy'' $H_X = - \int p_X(x) \ln p_X(x) \, dx$
is not an intrinsic quantity of the phenomena concerning $X$. 
With a different parametr{is}ation, i.e.\ using the coordinates $y=f(x)$ with an invertible function $f$,
rather than $x$, the entropy of the same phenomenon would now be given by
$$H_Y = - \int p_Y(y) \ln p_Y(y) \, dy
$$ 
with $p_Y(y)=p_X(f^{-1}(y))/|f'(x=f^{-1}(y)|$. Therefore, one has 
$$
H_Y=H_X + \int p_X(x) \ln|f'(x)| \, dx
$$
The  MEP gives different solutions if different variables are adopted to describe the very
same phenomenon.

In order to avoid this dependence on the choice of variables, Jaynes later proposed a more 
{sophisticated} version of the MEP, in terms of the relative entropy:
$$
{\tilde H}=-\int \rho({\bf x})
 \ln \Bigr[ {\rho({\bf x}) \over q({\bf x})} \Bigl] d{\bf x} \,\ ,
$$
where  $q$ is a known probability density. Of course, ${\tilde H}$ depends on $q$; but, at variance 
with the entropy, it does not depend on the chosen variables. On the other hand, one must decide how 
to select $q$, and this issue is equivalent to the problem of choosing the ``proper variables''.
Therefore, even this more elaborate method is non-predictive, and 
we see no reason to pursue {the MEP approach further} in the field of statistical mechanics.
The {interested reader} is referred to the existing extensive 
literature, e.g.~\cite{Jay_67,Uff_95}, for further details.

\section{Summary}
Let us {summarise} the fundamental concepts 
discussed above. The purpose of Statistical Mechanics is the understanding of macroscopic behaviour 
in terms of the properties of the microscopic constituents of matter. This has been
achieved adopting:
\begin{itemize}
\item the ergodic hypothesis
\item the Boltzmann's principle (\ref{klogw}) as a bridge law,
\item the huge number of degrees of freedom ($N\gg 1$)
\end{itemize}
However, strictly speaking, the ergodic hypothesis cannot be verified, except in a few 
exceptional cases, as evidenced by the FPU numerical experiment and by the KAM theorem. 
On the other hand, Khinchin's strong mathematical results showed that Eq.(\ref{EH}) 
holds in the $N \to \infty$ limit, for a physically relevant class of observables.
Then, the success of Statistical Mechanics in describing macroscopic systems may be attributed to the following
facts:
\begin{itemize}
\item
although ergodicity is not exactly verified in realistic models, it does hold in
a weak sense, which is sufficient for the purposes of physics;
\item
the bridge law $S = k \log W$ links the microscopic mechanical quantity $W$ with the
emerging thermodynamic quantity $S$, through the Boltzmann constant $k$.
We stress the fact that the relation (\ref{klogw}) is a fundamental assumption of the same nature
 (and importance) {as} Newton's principles for mechanics.
\item
macroscopic objects are made of very large numbers of microscopic constituents.
The number of particles in macroscopic bodies is {of the }order of the Avogadro's number ($N_A\approx 6.02~10^{23}$).
Boltzmann's constant $k=\mathcal{R}/N_A$, where $\mathcal{R}$ is the universal gas constant{ that} takes the 
mind-boggling value $k \approx 1.38~ 10^{-23} J/K$, is an 
astonishingly physically powerful element of the bridge law. Because it 
constitutes {a unit of entropy (energy divided by temperature)}, $k$ binds mechanics 
and thermodynamics together. The very small numerical value of $k$ measures the 
``distance'' between the microscopic world and the macroscopic world.
\item Chaos is not a fundamental ingredient for the validity
of statistical mechanics: the naive idea that chaos implies good statistical properties is 
inconsistent\cite{castiglione2008chaos,Liv_87}.
Sometimes, even in the absence of chaos on can have good agreement between the time averages and their 
values predicted by statistical mechanics; such a result is in agreement with Khinchin's 
ideas\cite{Liv_87}.
\item How large $N$ must be for the framework outlined above to properly describe physical systems is
hard to tell. No doubt, $N=N_A$ is largely sufficient for the vast majority of equilibrium phenomena, 
meaning phenomena on the scale of our daily life. Reducing the values of $N$, violations of the thermodynamic
behaviours are boud to be met, but when that happens depends very strongly on the systems and on the 
observables under consideration. One may have thermodynamic-like behaviours at the nanometric scale for certain 
observables, but not for others. A thumb rule is that fluctuations must be negligible compared to the averages,
but the onset of the non-thermodynamic behaviours is sitll largely to be understood. This is why the 
statistical physics of small systems is a thriving subject.
\end{itemize}

\bibliographystyle{unsrt}
\bibliography{biblio}

\begin{thebibliography}{10}

\bibitem{Skl_95}
L.~Sklar.
\newblock {\em Physics and chance}.
\newblock Cambridge University Press, 1995.

\bibitem{EmL_02}
G.~Emch and C.~Liu.
\newblock {\em The Logic of Thermo-Statistical Physics}.
\newblock Springer Verlag, 2002.

\bibitem{Pri_84}
I.~Prigogine and I.~Stengers.
\newblock {\em Order out of Chaos.}
\newblock Bantam Books, Toronto-New York-London-Sydney., 1984.

\bibitem{bricmont1995science}
J.~Bricmont.
\newblock Science of chaos or chaos in science?
\newblock {\em Annals of the New York Academy of Sciences}, 775(1):131, 1995.

\bibitem{MEP_2013}
S.~Press\'e, K.~Ghosh, J.~Lee, and K.~A. Dill.
\newblock Principles of maximum entropy and maximum caliber in statistical
  physics.
\newblock {\em Rev. Mod. Phys.}, 85(3):1115, 2013.

\bibitem{lebowitz1993boltzmann}
J.L. Lebowitz.
\newblock Boltzmann's entropy and time's arrow.
\newblock {\em Physics Today}, 46:32, 1993.

\bibitem{castiglione2008chaos}
P.~Castiglione, M.~Falcioni, A.~Lesne, and A.~Vulpiani.
\newblock {\em Chaos and coarse graining in statistical mechanics}.
\newblock Cambridge University Press, 2008.

\bibitem{Fer_56}
E.~Fermi.
\newblock {\em Thermodynamics}.
\newblock Dover Publications, New York, 1956.

\bibitem{Nag_79}
E.~Nagel.
\newblock {\em The structure of science}.
\newblock Hackett Publishing Company, 1979.

\bibitem{Ein_56}
A.~Einstein.
\newblock {\em Investigations on the Theory of the Brownian Movement}.
\newblock Dover Publications, 1956.

\bibitem{Cen_09}
M.~Cencini, F.~Cecconi, and A.~Vulpiani.
\newblock {\em Chaos: from simple models to complex systems}.
\newblock World Scientific, Singapore, 2009.

\bibitem{Khi_49}
A~I Khinchin.
\newblock {\em Mathematical Foundations of Statistical Mechanics}.
\newblock Dover Publications, 1949.

\bibitem{Maz_63}
P.~Mazur and J.~van~der Linden.
\newblock Asymptotic form of the structure function for real systems.
\newblock {\em Journal of Mathematical Physics}, 4:271, 1963.

\bibitem{Att_02}
P.~Attard.
\newblock {\em Thermodynamics and Statistical Mechanics}.
\newblock Academic Press, 2002.

\bibitem{Ma_85}
S.~K. Ma.
\newblock {\em Statistical Mechanics}.
\newblock World Scientific, Singapore, 1985.

\bibitem{Peliti}
L.~Peliti.
\newblock {\em Statistical Mechanics in a Nutshell}.
\newblock Princeton University Press, 2011.

\bibitem{Per_78}
A.~Peres.
\newblock Unperformed experiments have no results.
\newblock {\em Am. J. Phys}, 46(7):745, 1978.

\bibitem{Jay_67}
E.T. Jaynes.
\newblock Foundations of probability theory and statistical mechanics.
\newblock In M.~Bunge, editor, {\em Delaware Seminar in the Foundations of
  Physics}, volume~1, page~76. Springer-Verlag, Berlin, 1967.

\bibitem{Uff_95}
J.~Uffink.
\newblock Can the maximum entropy principle be explained as a consistency
  requirement?
\newblock {\em Studies in History and Philosophy of Science Part B: Studies in
  History and Philosophy of Modern Physics}, 26(3):223, 1995.

\bibitem{Liv_87}
R.~Livi, M.~Pettini, S.~Ruffo, and A.~Vulpiani.
\newblock Chaotic behavior in nonlinear hamiltonian systems and equilibrium
  statistical mechanics.
\newblock {\em Journal of Statistical Physics}, 48:539, 1987.

\end{thebibliography}

\end{document}